\documentclass[10pt,twocolumn,letterpaper]{article}

\usepackage[pagenumbers]{cvpr}      % To produce the REVIEW version
\usepackage{graphicx}
\usepackage{amsmath}
\usepackage{amssymb}
\usepackage{booktabs}

\usepackage[pagebackref,breaklinks,colorlinks]{hyperref}

\usepackage[capitalize]{cleveref}
\crefname{section}{Sec.}{Secs.}
\Crefname{section}{Section}{Sections}
\Crefname{table}{Table}{Tables}
\crefname{table}{Tab.}{Tabs.}

\def\cvprPaperID{55} % *** Enter the CVPR Paper ID here
\def\confName{CVPR}
\def\confYear{2022}

\begin{document}

\title{Conditional Vector Graphics Generation for Music Cover Images}

\author{Valeria Efimova\\
ITMO University\\
%Kronverksky Pr. 49, St. Petersburg, Russia\\
{\tt\small vefimova@itmo.ru}
\and
Ivan Jarsky\\
ITMO University\\
{\tt\small ivanjarsky@gmail.com}
\and
Ilya Bizyaev\\
ITMO University\\
{\tt\small  bizyaev@zoho.com}
\and
Andrey Filchenkov\\
ITMO University\\
{\tt\small afilchenkov@itmo.ru}\\
}
\maketitle

%%%%%%%%% ABSTRACT
\begin{abstract}
Generative Adversarial Networks (GAN) have motivated a rapid growth of the domain of computer image synthesis. 
As almost all the existing image synthesis algorithms consider an image as a pixel matrix, the high-resolution image synthesis is complicated.
A good alternative can be vector images. However, they belong to the highly sophisticated parametric space, which is a restriction for solving the task of synthesizing vector graphics by GANs.
In this paper, we consider a specific application domain that softens this restriction dramatically allowing the usage of vector image synthesis. 

Music cover images should meet the requirements of Internet streaming services and printing standards, which imply high resolution of graphic materials without any additional requirements on the content of such images. Existing music cover image generation services do not analyze tracks themselves; however, some services mostly consider only genre tags.
To generate music covers as vector images that reflect the music and consist of simple geometric objects, we suggest a GAN-based algorithm called CoverGAN. 
The assessment of resulting images is based on their correspondence to the music compared with AttnGAN and DALL-E text-to-image generation according to title or lyrics.
Moreover, the significance of the patterns found by CoverGAN has been evaluated in terms of the correspondence of the generated cover images to the musical tracks. 
Listeners evaluate the music covers generated by the proposed algorithm as quite satisfactory and corresponding to the tracks. Music cover images generation code and demo are available at \url{https://github.com/IzhanVarsky/CoverGAN}.
%Almost all the existing image synthesis algorithms consider an image as a pixel matrix, which is an obstacle for high-resolution image synthesis.
%Vector images could be a suitable alternative. However, the task of synthesizing vector graphics is not in scope due to the highly sophisticated parametric space, to which such images belong to.
%The resulting images are assessed by users who evaluated how well the generated images correspond to the music in comparison with AttnGAN text-to-image generation by lyrics. 
%The proposed algorithm generates music covers perceived by listeners as quite satisfactory and corresponding to the tracks. 
\end{abstract}

%%%%%%%%% BODY TEXT
% переделать картинку по шаблону
\begin{figure}[t] 
    \centering
    \includegraphics[width=0.9\columnwidth]{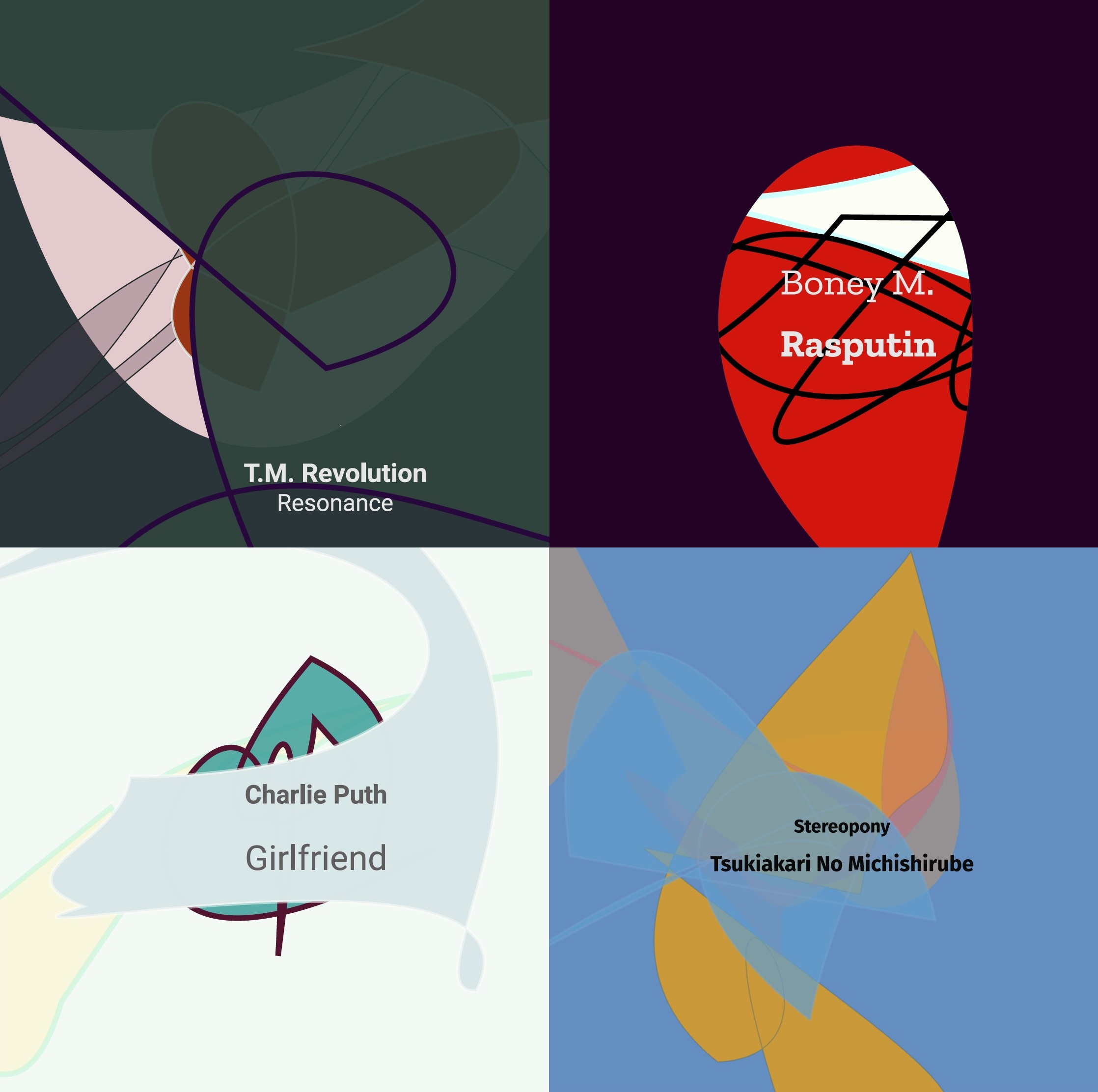} 
    \caption{Music cover images generated in vector format with the proposed CoverGAN.}
    \label{fig:res}
\end{figure}

% где-нибудь вначале точно должна быть ссылка на спецификацию svg: ~\cite{Cohn2000ScalableVG}

\section{Introduction}
%Image synthesis is a trending research direction due to a high demand in many fields due to drawing images is a time-consuming process. 
 Drawing images manually is a time-consuming process, therefore,  image synthesis is a trending research direction due to its high demand in many fields.    
Various Generative Adversarial Networks (GANs)~\cite{goodfellow2014generative}, Variational Autoencoders (VAEs)~\cite{kingma2019introduction}, Autoregressive Models~\cite{oord2016conditional}, and other models~\cite{bond2021deep} have been proposed to address this challenge. 
They all work with bitmap generation representing an image as a matrix of pixels. 
This results in limitations of the image quality when its resolution is meant to be high. 
Moreover, new networks capable of synthesizing high-resolution bitmap images are very time- and power-consuming. 

An alternative to bitmap images is vector graphics that may help avoid fuzzy lines and artifacts typical of GANs~\cite{hertzmann2020visual} and achieve sufficient image resolution. However, vector image generation from scratch has not been sufficiently studied. Recently, the vector images are generated containing only points or thin curves~\cite{clipdraw,li2020diffvg}, however artists usually draw vector images with color-filled shapes. 
%, and, to the best of our knowledge, no works are published on this topic.
% векторная картинка генерируется не из замкнутых закрашенных форм, а из точек или тонких кривых.

Being motivated by the potential of vector graphics synthesis and lack of relevant work, we formulate our research question as: {is it possible to synthesize visually appealing vector graphic images with machine learning methods?}

The main challenge here is that vector images are highly different from raster images. Raster images are represented as a two-dimensional rectangular matrix or grid of square pixels, while a common vector image is an XML~\cite{xml-docs} file containing descriptions of geometric lines, which can be completed or not. The most useful tag of the XML file is the \texttt{<path>} tag; it defines a set of rules for drawing straight segments, B\'ezier curves of the 2nd and 3rd order, and circular arcs. This tag mainly consists of control points specifying the type and shape of the rendered figure. 

%A promising application area is music cover images, which are a key component of music marketing, which, despite the digital distribution of music, has retained its importance: listeners usually navigate music collections by small cover images before actually listening to the tracks. 
A promising application area is music cover images, which have retained their importance as a key component of music marketing despite the digital music distribution. Listeners usually navigate music collections by small cover images before actual listening to tracks.
The attractiveness of cover image does not determine music quality but draws the attention of online music consumers to discographies~\cite{cook2013music}, and remains one of the most important issues of popular culture, especially on physical media~\cite{medel2014resurrection}.
Noteworthy, streaming services require high-resolution cover images (a common recommendation is $3000 \times 3000$ pixels).
Independent musicians order cover image production from artists, designers, and photographers, or make attempts to create them on their own, which, however, requires special skills musicians usually lack. 

Despite the recent advancements in text-to-image generation~\cite{xu2018attngan, zhang2017stackgan}, few audiovisual models have been developed. 
Existing models are mostly aimed at correlating sound information with certain real scenes~\cite{qian2020multiple}, actions~\cite{gao2020listen}, or actors~\cite{oh2019speech2face}. 
Currently, various simplified editors and template constructors exist, but there are only four publicly available Internet services offering musicians computer-generated images as cover images for their musical compositions: Rocklou Album Cover Generator~\cite{rocklou}, Automated Art~\cite{aa}, GAN Album Art~\cite{ganaa}, and ArtBreeder~\cite{artbreeder}.

We propose a GAN, which can generate vector graphics using only image supervision. Therefore, we suggest an approach to cover generation in vector format for musical compositions and call it CoverGAN. The samples of generated cover images can be seen in Fig.~\ref{fig:res}. 
%At the beginning, we would be able to create quite simple images and, as for future work, we plan to generate .

The structure of the paper is as follows. 
In Section~\ref{sec:work}, we briefly describe the results achieved.  
In Section~\ref{sec:method}, we present a novel method for the cover generation task. 
Experimental evaluation is presented in Section~\ref{sec:res}.
Limitations of the approach presented are discussed in Section~\ref{sec:limit}.
Section~\ref{sec:conclusion} concludes the paper and outlines future research.

\section{Related work}\label{sec:work}
The construction of an audiovisual generative model using musical composition includes the extraction of its sound characteristics and conditional image generation. Cover generating services have also already been created for musical compositions.

\subsection{Music information retrieval}
Music information retrieval (MIR) from audio data is a well-studied interdisciplinary field of research~\cite{mir-dl,schedl2014music,moffat2015evaluation} based on signal processing, psychoacoustics and musicology.
Currently, various information can be obtained automatically from audio data.
Librosa library~\cite{librosa} can extract mel-frequency cepstral coefficients (MFCC)~\cite{davis1980comparison}, which measure the timbre of a music piece and are often used as a feature for speech recognition. They are also widely used for classification based on acoustic events in the habitat. Essentia library~\cite{Bogdanov2013EssentiaAA, essentia-docs} offers methods for extracting tonality, chords, harmonies, melody, main pitch, beats per minute, rhythm, etc.
 
 %In addition, it should be noted that algorithms based on the study of spectrograms of musical works are also used. %Чет не смог найти таких
 %Also lot of dataset for music information retrieval exists \cite{http://mirlab.org/dataset/public/}.
 % TODO: добавить много ссылок и перечисление, какие фичи в принципе можно извлекать со ссылками 

\subsection{Conditional image generation}
Conditional image generation is the process of constructing images corresponding to specified criteria based on certain input data (most often categorical). 
The Conditional Generative Adversarial Network (cGAN)~\cite{mirza2014conditional} is one of the most popular model architecture applied. 
Unlike a common unconditional GAN, in this model the condition is passed to the input of both the generator and the discriminator. It becomes possible to generate an image based on a text condition. 
%One of such models is, for example, StackGAN~\cite{zhang2017stackgan}. In it, the image generation takes place in two stages, at the first stage a primitive image shape is created and the colors of objects are set, at the second stage defects of the previous stage are corrected and smaller details are added.

Successful modification of the cGAN model is AttnGAN~\cite{xu2018attngan}. This model considers the mechanism of attention~\cite{vaswani2017attention} as a learning factor, which allows selecting words to generate image fragments. Due to modifications, this network shows significantly better results than traditional GAN systems.
ObjGAN~\cite{Li2019ObjectDrivenTS} also uses the attention. However, its basic principle of image generation is to recognize and create individual objects from a given text description.
MirrorGAN paper~\cite{mirrorgan} uses the idea of learning by redescription and consists of three modules: the semantic text embedding module%(STEM)
, global-local collaborative attentive module for cascaded image generation%(GLAM)
, and semantic text regeneration and alignment module.% (STREAM).

Not only cGANs can generate images by condition. 
A well-known model is Variational Autoencoder (VAE)~\cite{kingma2019introduction}. A striking representative of this approach is the NVAE model~\cite{Vahdat2020NVAEAD}, which uses depth-wise separable convolutions, residual parameterization of Normal distributions, and spectral regularization that stabilizes training.

A well-known project from OpenAI~\cite{openai} is the DALL-E model~\cite{dall-e}, which is a decoder-only transformer~\cite{vaswani2017attention} built on GPT-3~\cite{gpt3} architecture and capable of generating realistic images based on provided text condition.
At the same time, OpenAI has released a Contrastive Language–Image Pre-training (CLIP) model~\cite{clip} that learns the relationship between the whole sentence and the image it describes, and can act as a classifier.

% Возможно у меня не очень по шаблону получилось:
\begin{figure*}
  \hfill
  \begin{subfigure}{0.2\linewidth}
    %\fbox{\rule{0pt}{2in} \rule{.9\linewidth}{0pt}}
    \includegraphics[width=\linewidth]{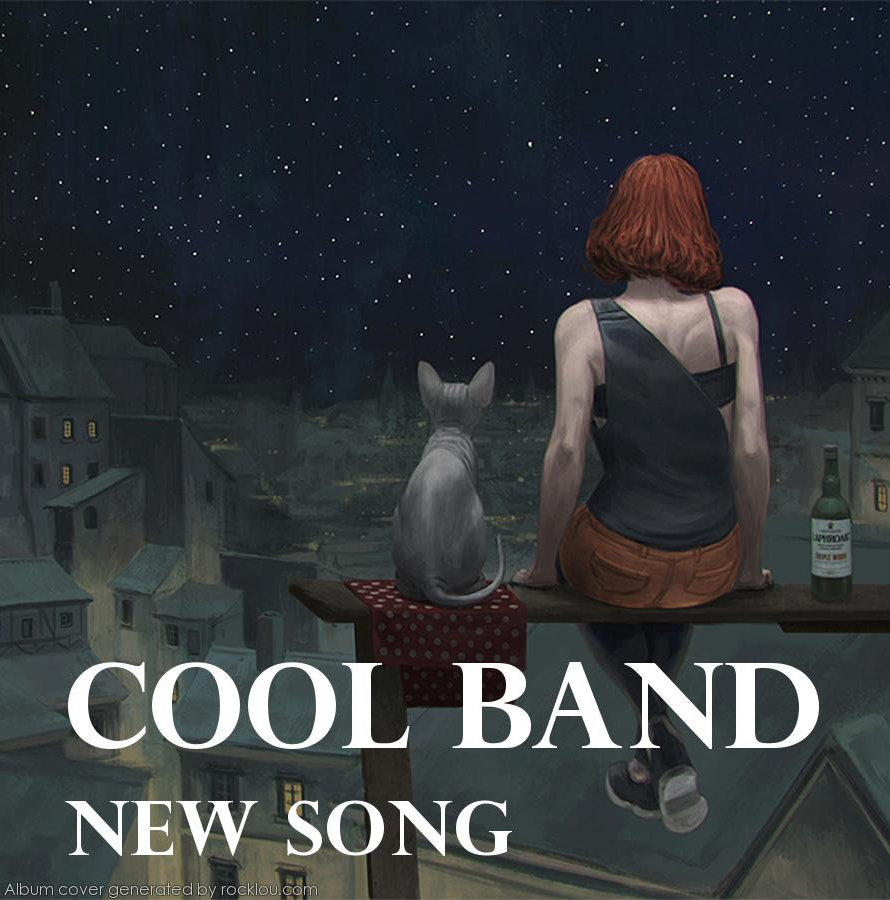}
    \caption{Rocklou Cover Generator}
    \label{fig:short-a}
  \end{subfigure}
  \hfill
  \begin{subfigure}{0.2\linewidth}
    %\fbox{\rule{0pt}{2in} \rule{.9\linewidth}{0pt}}
    \includegraphics[width=\linewidth]{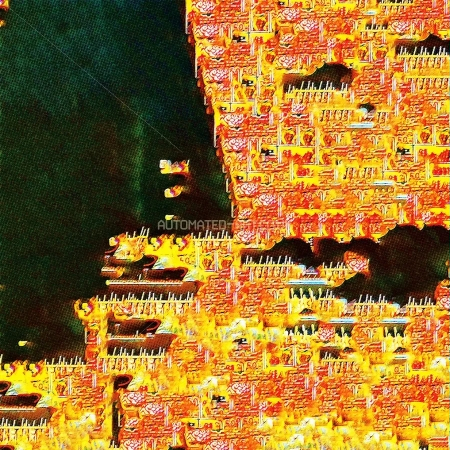}
    \caption{The Automated Art}
    \label{fig:short-b}
  \end{subfigure}
  \hfill
  \begin{subfigure}{0.2\linewidth}
    %\fbox{\rule{0pt}{2in} \rule{.9\linewidth}{0pt}}
    \includegraphics[width=\linewidth]{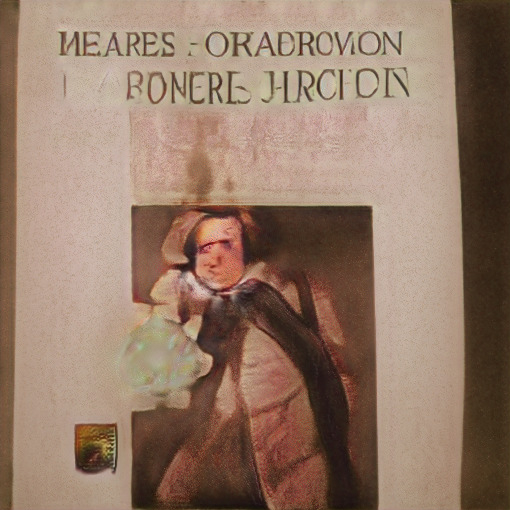}
    \caption{GAN Album Art}
    \label{fig:short-c}
  \end{subfigure}
  \hfill
  \begin{subfigure}{0.2\linewidth}
    %\fbox{\rule{0pt}{2in} \rule{.9\linewidth}{0pt}}
    \includegraphics[width=\linewidth]{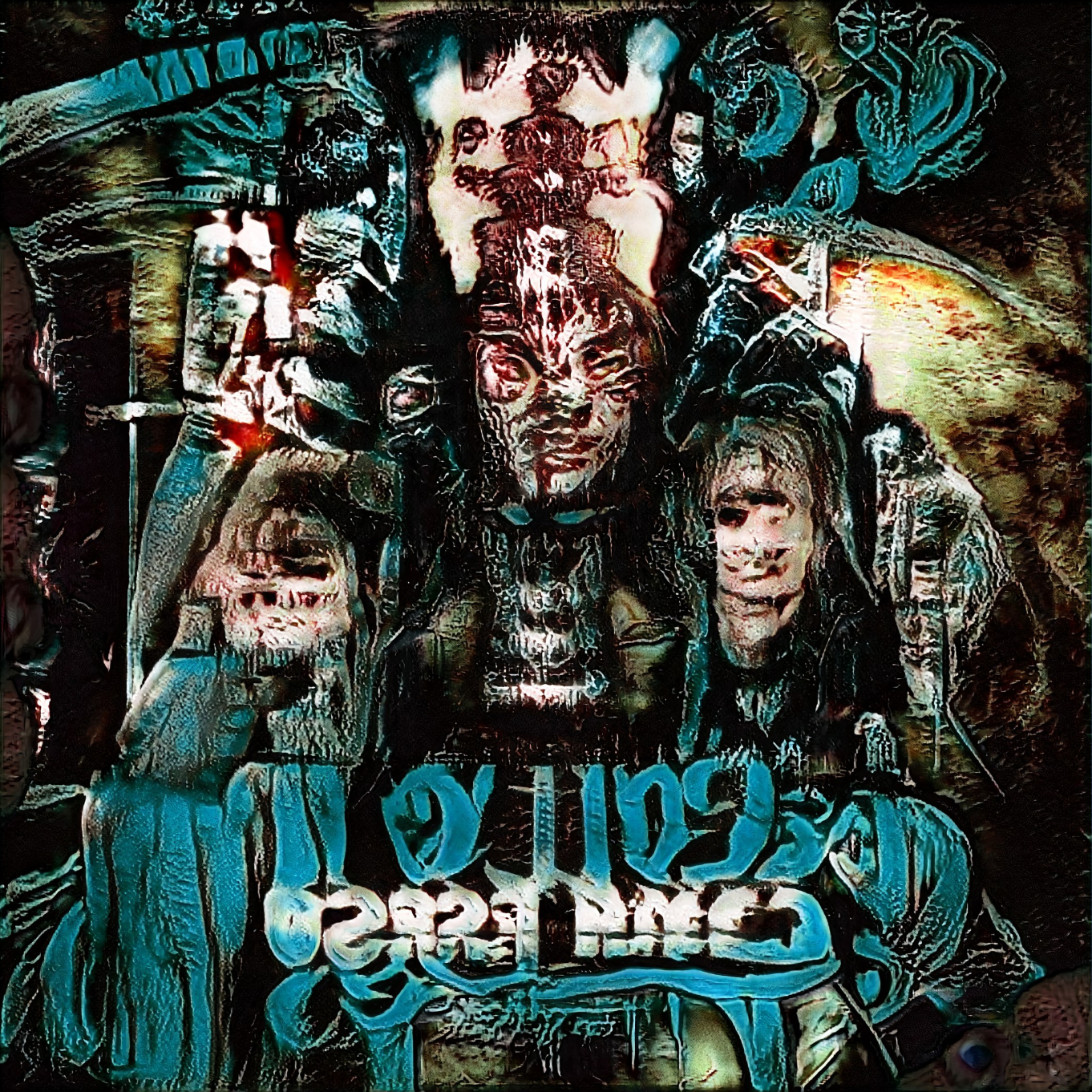}
    \caption{ArtBreeder}
    \label{fig:short-d}
  \end{subfigure}
  \caption{Automated cover generation services examples.}
  \label{fig:services}
\end{figure*}

\subsection{Vector images generation}
The generation of vector images, in contrast to the generation of raster images, is still poorly studied. However, several papers on this topic have been recently published.

%At the very beginning of studying the generation of vector images, there were works that transformed vector images to raster images and conducted training on raster images. However, it quickly became clear that this method would not be very successful. % нужно найти и вставить ссылки на такие работы (я надеюсь, что то, что я написал правда и это так и было =) ).
%Soon more successful works were suggested.

The well-known SVG-VAE~\cite{svg-vae} and DeepSVG~\cite{deep-svg} models are capable of generating vector images. However, they require vector supervision and need collecting large datasets of vector images, which is complicated.

DiffVG~\cite{li2020diffvg} has become a great achievement in creating vector images. This work proposed a differentiable rasterizer for vector graphics, which bridges the raster and vector domains through backpropagation and enables gradient-based optimization. This method supports polynomial and rational curves, stroking, transparency, occlusion, and gradient fills.

In the paper~\cite{reddy2020discovering} the problem of differentiable image compositing was considered by proposing the DiffComp model. The authors presented a differentiable function, which composes provided discrete elements into a pattern image. Using this operator, vector graphics becomes connected with image-based losses, and it becomes possible to optimize provided elements in accordance with losses on the composited image. 
% про всякую генерацию вектора (http://geometry.cs.ucl.ac.uk/projects/2020/diffcompositing/ и еще как минимум 3 статьи 2019-2021 гг)

Another successful work is Im2Vec~\cite{reddy2021im2vec}, which discusses  synthesizing vector graphics without vector supervision. Inspired by ideas of SVG-VAE~\cite{svg-vae} and DeepSVG~\cite{deep-svg}, authors proposed an end-to-end VAE that encodes a raster image to a latent code and then decodes it into a set of ordered closed vector paths. After that, these paths are rasterized and composited together using the  DiffVG and DiffComp solutions mentioned above. An important element is the proposed path decoder, which is capable of decoding the latent code into closed B\'ezier curves. It becomes possible due to sampling the path control points uniformly on the unit circle, which is then deformed and transformed into final points in the absolute coordinate system of the drawing canvas. In this paper multi-resolution raster loss solves the problem when at the early stages the gradients have a small area of influence using rasterisation at multiple resolutions. The authors claim that Im2Vec shows better results than SVG-VAE and DeepSVG.

ClipDRAW~\cite{clipdraw} paper is devoted to generating vector images based on the input text. This model combines CLIP language-image encoder and DiffVG rasterizer. Initially, a set of random B\'ezier curves is generated, after that, they gradually transform into understandable silhouettes. Also, the model allows creating more or less realistic pictures depending on a given number of strokes.

\subsection{Approaches to the automatic creation of a cover for a musical composition}
\label{subsec:already-created-services}
%Our model generates vector covers for musical compositions. 
For the best of our knowledge, four services have been proposed to generate covers for musical compositions. The examples of the generated covers for each of them are presented in Fig.~\ref{fig:services}.

The Rocklou Album Cover Generator~\cite{rocklou} service allows generating covers with resolution of $900 \times 900$ pixels specifying the name of the artist and the title of the track, and selecting the genre of the audio track. To generate the album covers, it picks one of $160$ fonts and one of $1500$ template images. However, the covers are only for inspiration, they are not allowed to be used commercially. 

The Automated Art~\cite{aa} service allows selecting from pre-generated covers using GAN, but they contain a lot of fuzzy lines and artifacts. Two types of licensing are provided: One Time Use and Extended. Both of them limit the cover resolution to a maximum of $480000$ pixels in total, permit to use the purchased media in one project only and contain many other prohibitions in use. 

The GAN Album Art~\cite{ganaa} website displays a randomly selected low-resolution image from pre-generated covers, with an internal division by genre, but without any input data and without specifying a license. In addition, the generated covers contain a lot of characters from different and seemingly non-existent languages, which are impossible to read and understand.

The ArtBreeder~\cite{artbreeder} service allows creating up to $5$ high-resolution images per month for free based on user-defined color preferences and random noise. All generated images are considered the public domain. However, the resulting covers contain fuzzy figures, artifacts, as well as fragments of signatures and human bodies.

In addition to the Internet services listed above, the use of GAN for generating covers of musical compositions is found in the paper~\cite{hepburn2017album}. The authors report successful generation of images with a resolution of $64 \times 64$ pixels for the specified genre labels and with genre-specific visual features. 
It can be noted that all of these works have significant drawbacks, such as the use of generation methods with a poor diversity of results, low resolution of output images, and the presence of a large number of artifacts. Moreover, they do not analyze a music track itself. 
As seen from all literature reviewed, all the approaches did not solve the task to generate music cover images of acceptable quality and corresponding to music track. Therefore, we have challenged ourselves to develop such a model.
%Our initial task of creating music cover images of acceptable quality and corresponding to music track remains unresolved.

\section{Method}\label{sec:method}
The creative nature of the problem has motivated us to use unsupervised learning, and we apply the Conditional Generative Adversarial Network (cGAN)~\cite{mirza2014conditional}. 
%In this work, much interest is paid to generating cover images based on some sound features. 
In this work, music cover images are generated based on several sound features, which is a hallmark of this research. 
Both generator and discriminator inputs are conditioned. The condition can be an embedding of the entire music track or its fragment~\cite{duarte2019wav2pix}, as well as some additional data, for instance, a track emotion indicated by a musician. 
It is required to train the mapping both from an audio embedding and a random vector to an output vector to find a correlation of the cover image content with the features obtained from audio data. 
%Both generator and discriminator inputs are conditioned, and it is required to train the mapping both from an audio embedding and a random vector to an output vector, which should find a correlation of the cover image content with the features obtained from audio data. 

\subsection{Input features}
% Было:
% We have used several common algorithms for calculating the following music features~\cite{Bogdanov2013EssentiaAA}:
% \begin{itemize}
% \item MFCC~\cite{davis1980comparison}, 
% \item spectral contrast~\cite{Jiang2002MusicTC}, 
% \item spectral peaks~\cite{peaks}, 
% \item loudness~\cite{Skovenborg2004LoudnessAO},
% \item danceability~\cite{streich2005detrended}, 
% \item beats per minute and onset rate, 
% \item mean beats volume, 
% \item Chromagram~\cite{Korzeniowski2016FeatureLF}, 
% \item music key and its scale. 
% \end{itemize}
% Стало:
We have used several common algorithms for calculating the following music features~\cite{Bogdanov2013EssentiaAA}: MFCC~\cite{davis1980comparison}, spectral contrast~\cite{Jiang2002MusicTC}, spectral peaks~\cite{peaks}, loudness~\cite{Skovenborg2004LoudnessAO}, danceability~\cite{streich2005detrended}, beats per minute and onset rate, mean beats volume, Chromagram~\cite{Korzeniowski2016FeatureLF}, music key and its scale.
% Ответ рефери: "This work uses quite a lots of audio features. However, the necessity of doing so is not clearly shown." Думаю лучше хотя бы в общих чертах пояснить. Например:
These parameters can strongly influence musical perception~\cite{Color-Music-Association,freeman2020musical,lindborg2015colour,tsiounta2013classical}. 
For example, covers for songs with a rough voice, high volume, non-standard alarming fast rhythm or tonality with abundant musical chromaticisms, usually contain dark shades and sharp lines. At the same time, for the quiet, calm, and harmonious tracks the light and soft colors are prevailing on the covers.

The track is resampled to a frequency of $44100$Hz. After that music features are calculated for $10$-second track fragments with $5$-second overlapping and then normalized. 

Emotions in musical compositions are one-hot encoded as a vector with the length equal to the total number of emotions. The positions corresponding to the selected emotions are encoded with ones, and the rest are encoded with zeros. 

\subsection{CoverGAN}
\begin{figure}[t]
  \centering
  \includegraphics[width=0.9\linewidth]{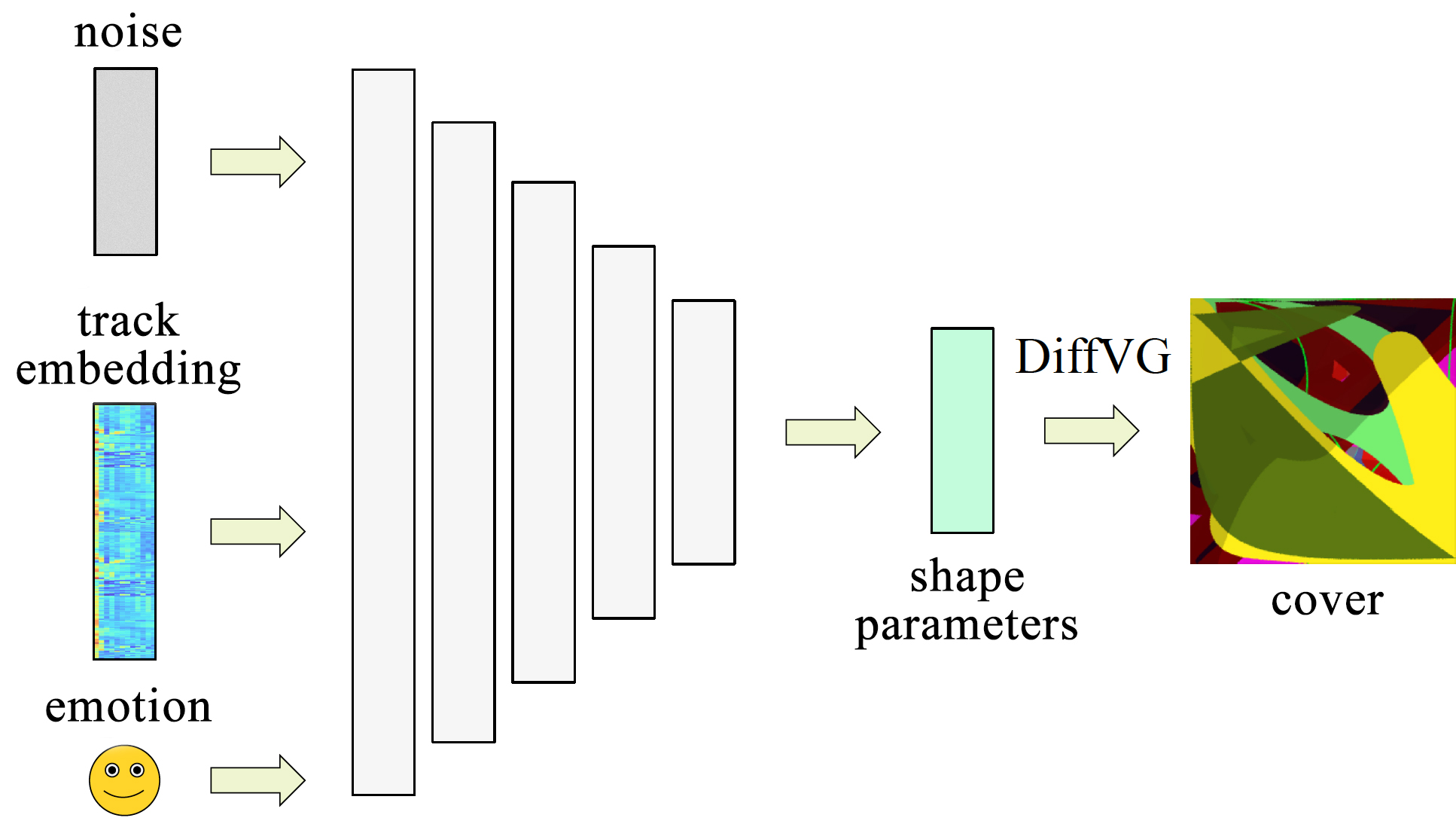}
   \caption{Fully-connected generator architecture.}
   \label{fig:lingen}
\end{figure}

The noise vector, encoded track emotion selected by a musician and audio features are passed as an input to the generator that creates a description of an image consisting of vector primitives. 
% Было:
% We suggest two architectures of the generator.
% Стало:
We have tested two architectures of the generator.
The first one is presented in Fig.~\ref{fig:lingen}. It consists of $5$ fully-connected layers with LeakyReLU with slope $0.2$ for $4$ layers and sigmoid activation function for the last one. As the result, output range is $(0; 1)$, which allows direct setting of color channels values. The coordinates of the points are set relative to the size of the canvas. However, the actual canvas size available to the generator for each of the dimensions is twice as large as the visible one; this gives the generator the ability to place parts of shapes outside the visible area. The thickness of the outlines is predicted relative to the specified maximum. To prevent internal covariate shift and speed up learning, batch normalization with a momentum of $0.1$ is applied.

% Было:
% The second architecture of the generator is presented in Fig.~\ref{fig:rnn}. It consists of a three-layer recurrent neural network with long short-term memory (LSTM) and several fully-connected layers corresponding to the optimized shape parameters: point coordinates, outline thickness, transparency, fill color, and outline color.
% The initial cell and latency values of the LSTM of $256$ features are determined by a fully-connected layer based on the emotions and noise vectors. Embeddings of short fragments of the audio file are sequentially fed to the model input; copies of the vectors of noise and moods are also added to each input. The LSTM output from each fragment is transferred to fully-connected layers that define the parameters of the shapes. For all output layers, except for the coordinate prediction, the sigmoid activation function is used, for the coordinates, the hyperbolic tangent activation function is used. 
% Стало:
The second tested architecture of the generator, consisted of a three-layer recurrent neural network with long short-term memory (LSTM) and several fully-connected layers corresponding to the optimized shape parameters: point coordinates, outline thickness, transparency, fill color, and outline color. Although this model has not yet brought an acceptable result, work on its refinement and improvement continues.

% \begin{figure}[t]
%   \centering
%   \includegraphics[width=0.9\linewidth]{img/rnn_eng.jpg}
%   \caption{Recurrent generator architecture.}
%   \label{fig:rnn}
% \end{figure}

For correct error backpropagation during model training, the rasterization of vector cover images must be differentiable, which is achieved by using the diffvg library~\cite{li2020diffvg} as the renderer. 
However, diffvg library cannot optimize the topology, such as adding and removing shapes, changing their order and type; therefore, the number of shapes is fixed to $3$ cubic B\'ezier curves of $4$ segments. Each B\'ezier curve consists of $13$ points: one initial and $3$ points by segment. 
We also create a square to represent the canvas color.
For the first model the number of curves is fixed and equal to $3$; for the second architecture, it corresponds to the number of embeddings of track fragments received by the generator.

\begin{figure}[t]
  \centering
  \includegraphics[width=0.9\linewidth]{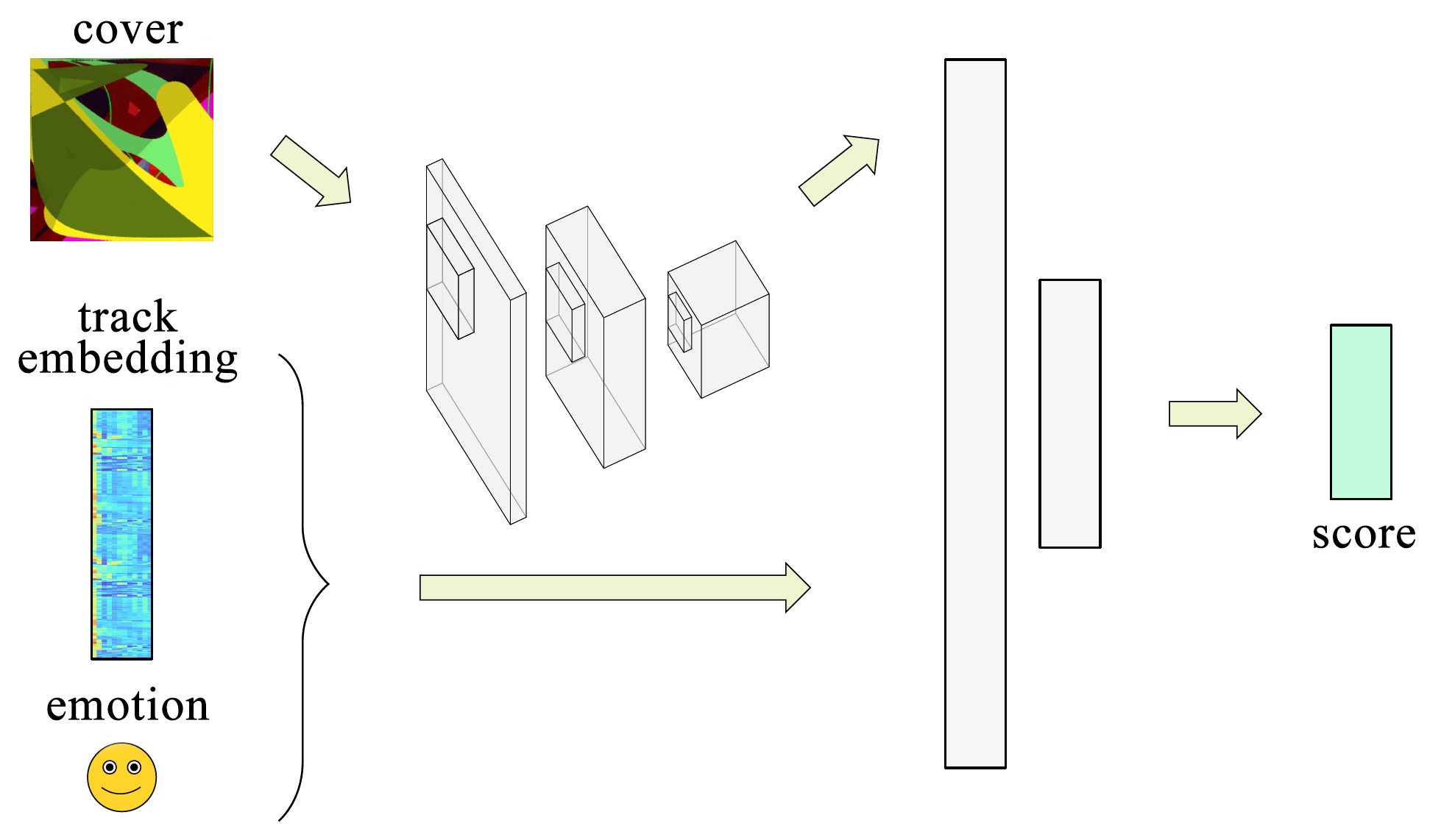}
   \caption{Discriminator architecture.}
   \label{fig:discr}
\end{figure}

The discriminator is presented in Fig.~\ref{fig:discr}. 
The model consists of $3$ convolutional and $2$ fully-connected layers. Tensors of real and generated covers with $3$ color channels (RGB) are provided as the input to the first convolutional layer. Real and generated covers are Gaussian blurred to prevent the influence of real covers noise on the discriminator decision.
%Tensors of real and generated covers with $3$ color channels (RGB), which are Gaussian blurred to prevent the influence of noise on real covers on decision-making, are provided as the 
%input to the first convolutional layer. 
Then, the first layer outputs $24$ channels, the subsequent ones {{--}} $48$ each. The first fully-connected layer takes as input the result of convolution, the mood vector and embedding of a large fragment of the musical composition. The number of features of the next layer is $64$ times less. The output of the network is one number {{--}} the assessment of the realism of the cover, which takes into account the compliance with audio data. LeakyReLU is the activation function on all layers, except the last one. Its slope for convolutional layers is $0.1$, for fully-connected layers {{--}} $0.2$. On the last layer, in accordance with the recommendation of the Wasserstein GAN (WGAN)~\cite{arjovsky2017wasserstein}, the activation function is not applied. Layer normalization and normalization by elimination with a probability of $0.2$ are used.

The first loss function of the discriminator is the Wasserstein loss with the gradient penalty~\cite{gulrajani2017improved}. 
Additional stimulation of the discriminator training in the correspondence of covers to the characteristics of musical compositions is provided by the secondary loss function. It is set as the difference of the average scores on the rearranged cyclic shift and on the corresponding covers from the training set batches, as follows (see Eq.~\ref{eq:1}):
% \begin{equation}
%   L_1 = D(G_{\theta}(z, a, e), a, e)) - D(r, a, e) + gp * c_{\lambda}
%     \label{eq:1}
% \end{equation}
\begin{equation}  
  L_2 = D(\tilde r, a, e) - D(r, a, e),
  \label{eq:1}
\end{equation}
where $a$ is a batch of audio embeddings, $e$ is a batch of emotion embeddings, $r$ is a batch of real cover images, $\tilde r$ is a random cyclic shift of the real covers $r$. The cyclic shift is important, because we do not want at least one cover in a batch to remain in its original place as it might with random shuffling.

% О чем этот второй лосс (судя по коду): у нас есть батч с реальными обложками + реальными муз.фичами + настроениями. Делаем из этого батча новый, у которого тензор реальных обложек заменим на его рандомный циклический сдвиг, а муз.фичи и настроения оставляем такими же. Берем средние оценки дискриминатора по первому батчу и средние оценки по второму и из второго значения вычитаем первое. Это и есть второй лосс. Честно говоря, не очень непонятно, как и почему он нужен и работает.

We considered the possibility of shape parameters selection in a separate (nested) GAN. Then, the external GAN would operate with vectors of the latent space of the internal one. However, despite the application of the pre-training on a Variational Autoencoder (VAE) to stabilize the training of the internal GAN, this approach led to no results. Currently, the nested GAN is not used.

After captioning (see Subsec.~\ref{subsec:cap}), the resulting description of the cover is saved in a common vector graphics format (SVG)~\cite{Cohn2000ScalableVG} or rasterized. 
The generated image of the cover does not violate the rights of the authors of the covers used while training~\cite{gillotte2019copyright}. After generation, it can be licensed to the performer of the track.

\subsection{Captioning}\label{subsec:cap}
\begin{figure}[t]
  \centering
  \includegraphics[width=0.9\linewidth]{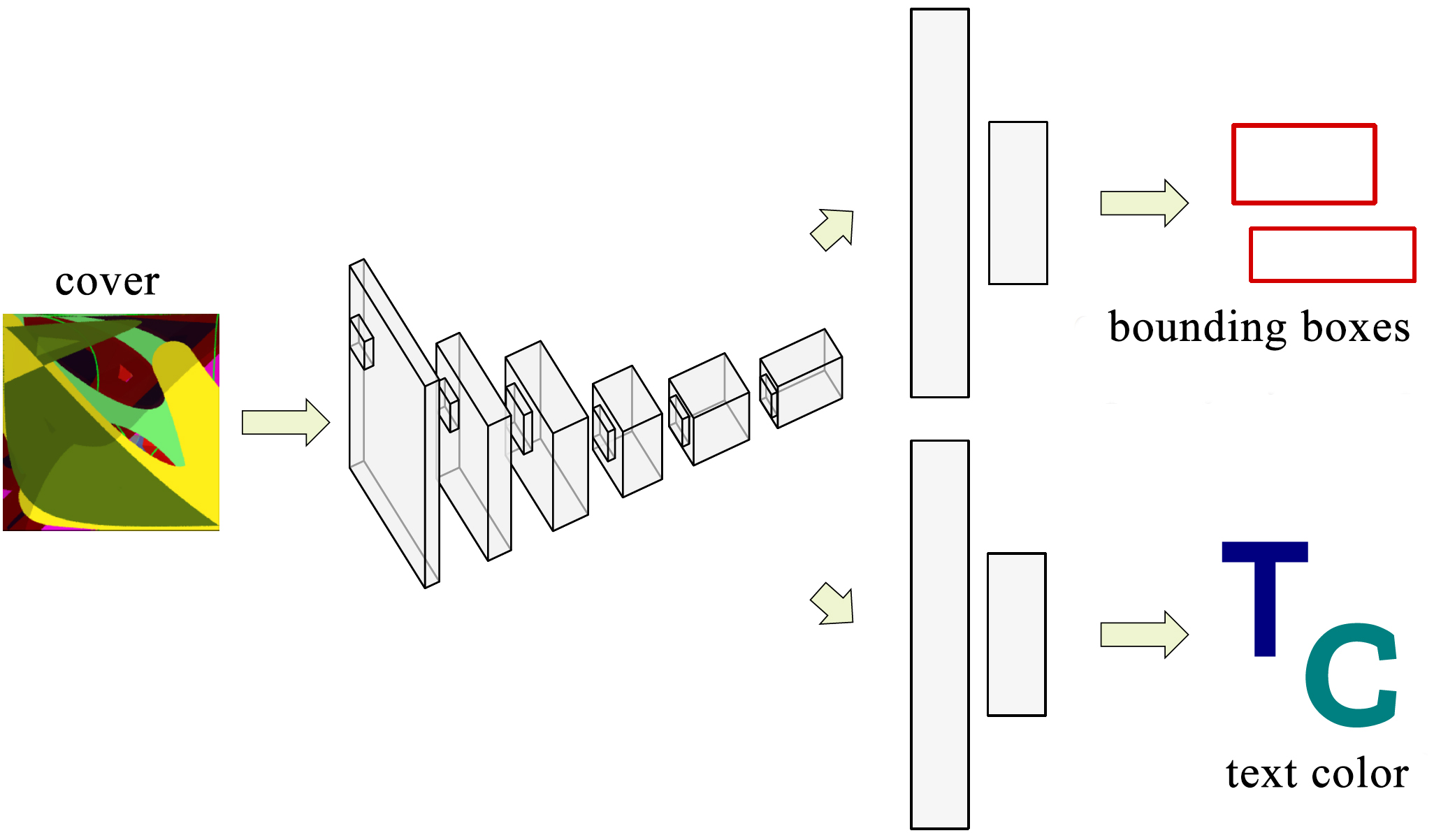}
   \caption{Captioning network architecture.}
   \label{fig:caption}
\end{figure}

To format generated images as complete covers, we add captions with authorship information, which optimal placement, color and style are determined by an additional neural network presented in Fig.~\ref{fig:caption}. It consists of $6$ fully-connected layers with $3 \times 3$ and $5 \times 5$ convolutional kernels and two $2$-layered sub networks, all with LeakyReLU activation with slope $0.01$. 

Images with four channels are fed to the network input, namely, three color channels (RGB) and the boundaries of the figures in these images, determined by the Canny edge detector~\cite{canny1986computational}. The network outputs text bounding boxes and text colors for captions. The optimal font size is selected accounting the predicted bounding rectangles. The font family is currently selected randomly from Google Fonts.

% Один из судей, не понял, как составлялся датасет, наверное нужно подробнее описать это:
The training dataset is described in details in section~\ref{sec:data}, it consists of original covers with text captions removed using a graphic editor.
%Edited cover images from a previously collected set of musical compositions (i.e. covers without title and artist) were used as a dataset for training the captioning network, for which the 
For training the captioning network, each image was labelled with text bounding boxes and colors. 
The labelling was partially automated by comparing the edited covers with the original ones. 

\section{Dataset}\label{sec:data}
The advantages provided by vector graphics in the field of scalability come at the cost of disadvantages: not every bitmap image can be rationally represented as a set of vector primitives. Thus, if the cover, for example, is a detailed photo, an attempt to vectorize and enlarge such a cover leads to a noticeable change in style. In this regard, it is reasonable to make a training dataset from covers that can be reproduced by a generator that draws vector primitives (basic geometric shapes and curves with strokes and fills). In addition, to better match the covers to musical compositions, it is preferable to collect a dataset from singles or mini-albums, where the album cover obviously refers to one of the tracks.

These limitations determine the need to collect a new dataset. In accordance with the listed requirements, a set of $1500$ musical compositions of various genres was collected. With the help of the Adobe Photoshop graphic editor~\cite{photoshop}, the captions were removed from the covers. On the crowdsourcing platform Yandex Toloka~\cite{toloka} the tracks were labeled with emotions  and included in the dataset. The label contains $2$-$3$ emotions of a musical composition from the following list: comfortable, happy, inspirational, joy, lonely, funny, nostalgic, passionate, quiet, relaxed, romantic, sadness, soulful, sweet, serious, anger, wary, surprise, fear. We can not release the dataset with music cover images due to copyright; however, its part containing $1500$ objects in the following format: track author {{--}} track title, $2$-$3$ emotions from the list, is publicly available\footnote{\url{https://www.kaggle.com/datasets/viacheslavshalamov/music-emotions}}.  

Furthermore, to obtain additional information about tracks (genres, release date, artist attributes, and others) was prepared a program that extracts available metadata directly from audio files and supplements them with information from public information databases (Wikidata~\cite{wikidata} and MusicBrainz~\cite{musicbrainz}). Track metadata is not currently used by the algorithm, but it is possible to do in the future.

After all, the collected dataset contains a number of covers that is not large enough to train deep learning models. Thus, to enlarge the resulting dataset, augmentation was applied; operations such as horizontal flip and $90$ degrees clockwise and counterclockwise rotation were applied to the covers. 

\section{Results}\label{sec:res}
Unlike the existing solutions, which generates covers that do not take into account musical compositions, the suggested model allows using extracted sound features in a generative algorithm. The created covers contain titles with information about authorship and track name and are available in a common vector graphics format without additional licensing restrictions on components. This allows us to speak about their compliance with generally accepted requirements for the design of musical compositions releases. 

\subsection{Training}
Training of neural networks was performed on the capacities of the Google Colaboratory~\cite{google-colab} service using the CUDA computing architecture. To achieve stable training of the model, various hyperparameters were tried, including different network architectures, the number of layers in models, learning rate, optimization algorithms, gradient penalty coefficient and the number of steps to update discriminator. The final version uses the generator consisting of fully-connected layers, a learning rate coefficient of $0.0005$ with a batch size of $64$, a five-time repetition of the discriminator training and a canvas size of $128 \times 128$ pixels. The training took $7200$ epochs based on a manual assessment of the quality of the generated images.

The Adam optimization algorithm~\cite{Kingma2015AdamAM} was chosen for CoverGAN training. Its gradients decay rate control coefficient $\beta_1$ was set to $0.9$ and for the second moments of gradients $\beta_2 = 0.999$. 

To train the captioning network, batch normalization with a following hyperparameters is used: a momentum of $0.1$ and normalization by elimination with a probability of $0.2$. The Adam algorithm with gradients decay rate control coefficient $\beta_1 = 0.5$ and for the second moments of gradients $\beta_2 = 0.999$ is chosen as the optimization algorithm.

During the training of the captioning network, $256 \times 256$ pixel images and a batch size of $64$ are used to select the design of signatures. The average quadratic error is used as a loss function. For validation, the Generalized Intersection over Union (GIoU)~\cite{Rezatofighi2019GeneralizedIO} metric is calculated and averaged for the entire dataset. 
In order to maximize it, training lasts $138$ epochs. The value of GIoU metric of $0.65$ has been achieved for the arrangement of rectangles; the average error value for colors on the entire training set reaches about $0.00014$ at the end of training.

\subsection{Comparison with text-to-image generation}
To assess the effectiveness of the proposed algorithm, we tested it on musical compositions not included in the training set. 
As a result, for different musical compositions, the algorithm created significantly different covers (Figure~\ref{fig:res}). 

Due to the lack of alternative algorithms that rely on musical features for cover creation, we used images generated by the AttnGAN~\cite{xu2018attngan} and DALL-E~\cite{rudalle} based on lyrics for the following quality comparison.
AttnGAN takes as input text of arbitrary length, when more modern approaches, such as DALL-E, focus on short texts (up to $200$ symbols) that are much shorter than a song lyrics. Thus, we also assess the quality of DALL-E-generated-images by query: 'Cover for the track \texttt{<title>} \texttt{<lyrics>}'.
% Также мы специально дообучили для генерации обложек уже существующую обученную на датасете COCO модель AttnGAN. В таблице она обозначена как `AttnGAN + covers`. Можно еще написать, что песни бывают на разных языках и мы переводили их на английский для обучения.
Furthermore, we fine-tuned AttnGAN model initially trained on COCO dataset on our dataset of vector covers. 

\begin{figure}[t] 
    \centering
    \includegraphics[width=1.0\columnwidth]{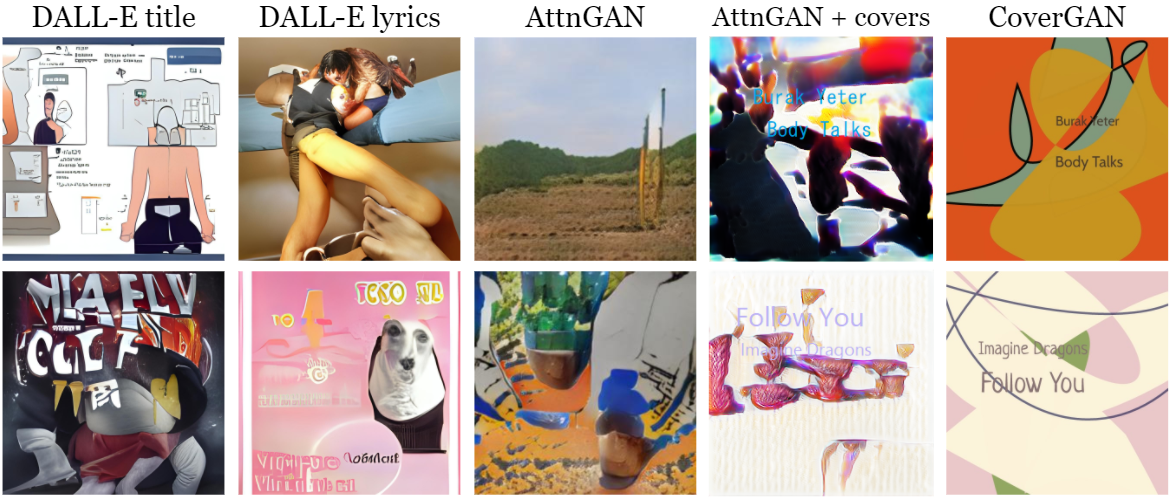} 
    \caption{Generated music cover images using different models; each row corresponds to real music track.}
    \label{fig:songs-comparing}
\end{figure}

We conducted a survey asking participants to listen to $15$ popular musical compositions and rate the covers generated by the proposed algorithm and AttnGAN on a scale of $1$ to $5$ ($1$ stands for completely inappropriate, $5$ stands for the perfect fit). $110$ assessors took part in the survey, $24$ of them identified themselves as musicians.
Normalized survey results are presented in Tab.~\ref{tab:attngan}. The examples of the generated images are shown in Fig.~\ref{fig:songs-comparing}. As it can be seen from the figure, DALL-E generates unknown symbols and undetermined shapes, AttnGAN generates patterns, whereas shapes generated by our CoverGAN are very simple. 

\begin{table}
  \centering
  \begin{tabular}{@{}lcc@{}}
    \toprule
    Method & All score & Musicians \\
    \midrule
    DALL-E$^t$ & $0.34 \pm 0.03$ & $0.31 \pm 0.05$ \\
    DALL-E$^l$ & $0.3 \pm 0.03$ & $0.26 \pm 0.05$ \\
    AttnGAN$^l$ & $0.44 \pm 0.04$ & $0.23 \pm 0.06$ \\
    AttnGAN$^l$ + covers & $0.36 \pm 0.03$ & $0.19 \pm 0.04$ \\
    CoverGAN (Ours) & $0.68 \pm 0.05$ & $0.71 \pm 0.05$ \\
    \bottomrule
  \end{tabular}
  \caption{
   Comparison of the covers generated by the proposed CoverGAN with DALL-E for titles denoted as DALL-E$^t$, DALL-E for lyrics denoted as DALL-E$^l$, AttnGAN$^l$ for lyrics, and AttnGAN for lyrics fine-tuned on covers denoted as AttnGAN$^t$ + covers. 
  'All score' column indicates the normalized score given by all assessors. 'Musicians' column indicates the normalized assessment score given by assessors identified themselves as musicians.}
  \label{tab:attngan}
\end{table}

The text-to-image generation models are very sensitive to the input text string, which often does not describe the entire track, and sometimes even contradicts the general mood of the song. Therefore, the resulting images do not often correspond to music. 

Although the AttnGAN$^t$ + covers model generates images with a monotonous background, the objects depicted have incomprehensible outlines and artifacts. 
Moreover, the generated images can have grid pattern.
These are probably the reasons that assessors rated this model worse than the previous one, as it can be seen from the table.

The CoverGAN score means that the generated cover images are quite satisfactory, however, further work in this direction is reasonable. 

\subsection{Different genres}
To assess the significance of the patterns found by the GAN model comparing sound and the generated covers, another survey was conducted using the crowdsourcing project Yandex Toloka. Its participants were asked to listen to $10$ musical compositions from various genre categories and choose the most suitable of the two generated covers for each. One of the covers was created by the generator directly based on the specified track, the other {{--}} based on a track from another category with the replacement of the caption text.

According to the results based on the assessments collected from $110$ listeners, the probability of the respondents, who have chosen the cover created by the generator for the track, was $0.76 \pm 0.03$. This score confirms the presence of significant differences between the proposed covers for listeners and indicates a moderate tendency of survey participants to agree with the conclusions of the generator. At the same time, there is a need for further improvement of the generative algorithm, including to increase the diversity of generated images.

\subsection{Qualitative Analysis}
To check the generalization ability of the CoverGAN, we analyzed the results of running the algorithm on musical compositions, which are not included in the training set. It was revealed that the algorithm creates significantly different covers for different tracks (see Fig.~\ref{fig:res}).
By changing the noise vector, we achieve the generation of alternative images to provide the user with a choice.

\begin{figure}[t] 
    \centering
    \includegraphics[width=0.8\columnwidth]{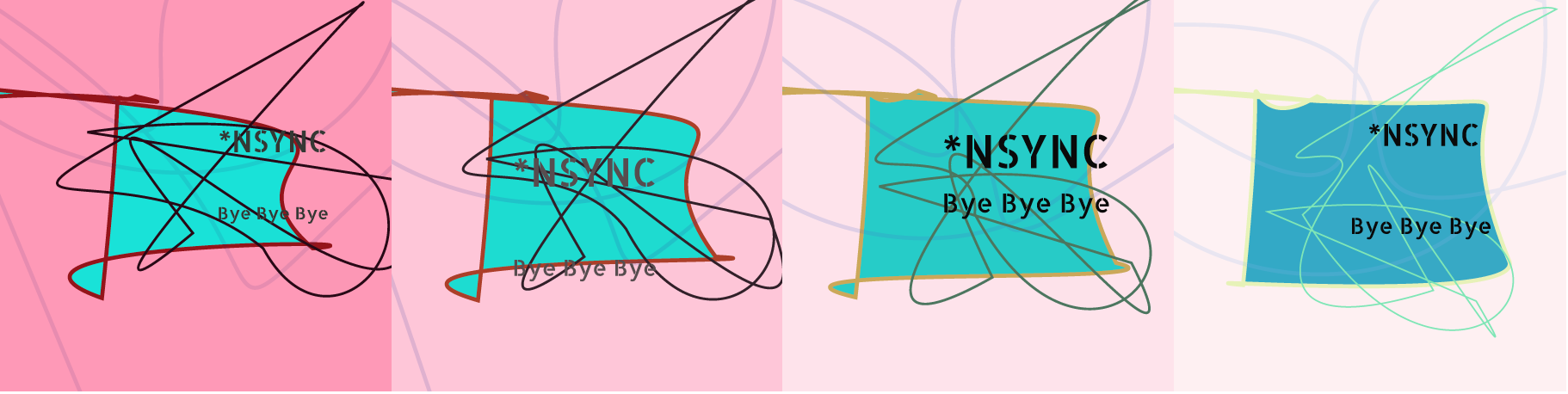} 
    \caption{The results of running the algorithm on several versions of the same musical composition with different volume levels (from left to right volume level: 100\%, 75\%, 50\%, 25\%).}
    \label{fig:volume-diff}
\end{figure}

For individual sound features, it can be traced how the generated cover image changes with a gradual change in the feature (for example, for the same musical composition at different volume levels as in Fig.~\ref{fig:volume-diff}). 

\begin{figure}[t] 
    \centering
    \includegraphics[width=0.8\columnwidth]{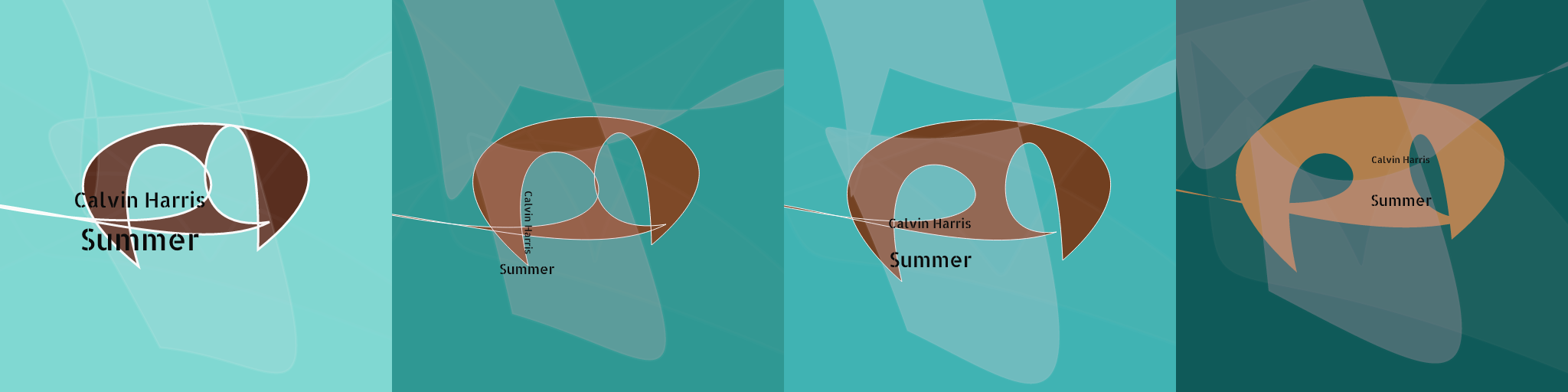} 
    \caption{The dependence of the cover on the specified emotion of the track (from left to right: comfortable, passionate, relaxed, wary).}
    \label{fig:emotion-diff}
\end{figure}

Moreover, user-specified emotions have a significant impact on the generated covers. Figure~\ref{fig:emotion-diff} shows the change of the cover when the specified emotion changes.

\begin{figure}[t] 
    \centering
    \includegraphics[width=0.6\columnwidth]{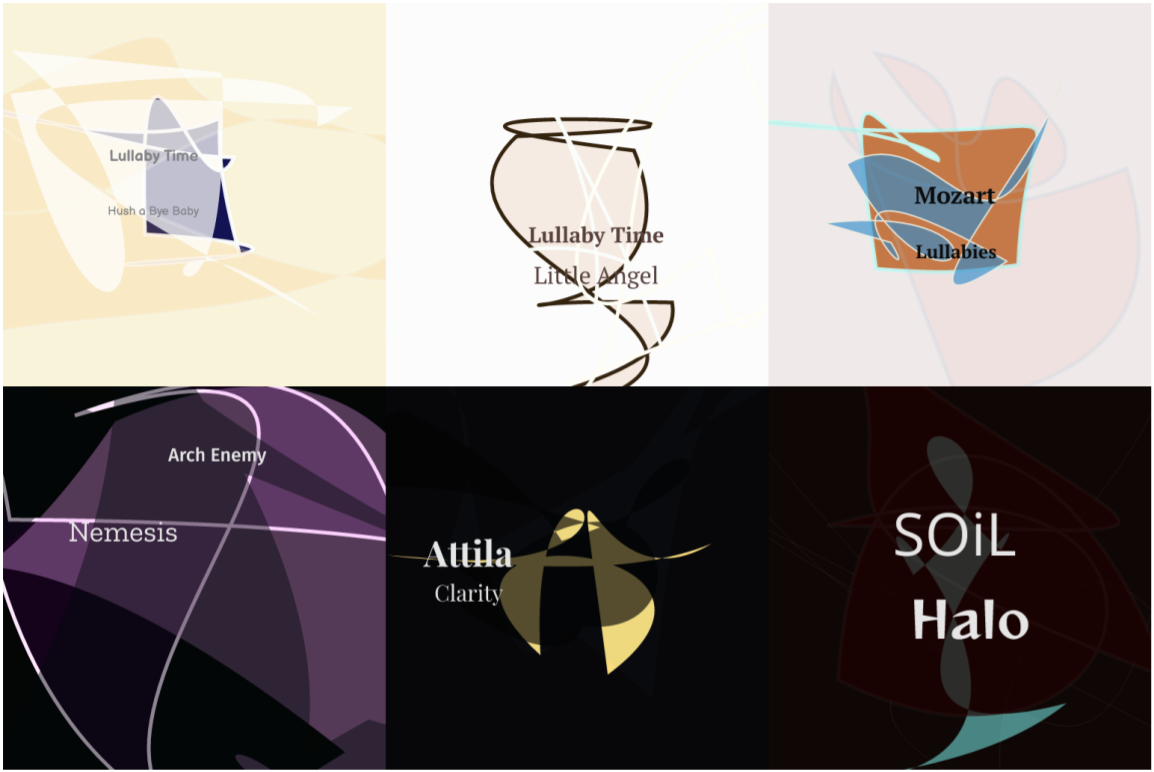} 
    \caption{Covers generated for lullabies and rock music.}
    \label{fig:lullaby-and-rock}
\end{figure}

Finally, on the test dataset, it was found that the covers of similar tracks also have some similarities. Therefore, for several lullabies, the algorithm suggested covers of light tones, while the covers generated for heavy metal genre %(a kind of rock music)
turned out to be dark (Fig.~\ref{fig:lullaby-and-rock}). Such connections are not universal, although their presence is consistent with the conclusions of the work~\cite{hepburn2017album}. In the future, such correspondences can be used to detect and analyze patterns found by a trained generative network.

\section{Limitations}\label{sec:limit}
The major limitation is the simplicity of generated paths. They usually do not have any %правильная геометрическая
regular shape such as a circle or a square. Moreover, it is difficult for CoverGAN to generate objects with visual semantics (i.e., person images, natural scenes). The generated images are abstract and cannot cover all scenarios in art design.
% Ответ одного из рефери: "I think the biggest limitation is that CoverGAN can only generate abstract images, which cannot cover all scenarios in art design."
% И еще один ответ: "It is difficult for CoverGAN to generate objects with visual semantics (i.e., person images, natural scenes, etc.) in the cover images."
% Еще можно добавить, что невозможно использовать градиенты в цветах, потому что diffVG в такое не умеет, хотя довольно распространненая вещь
% Добавил этот текст:
%In addition, CoverGAN can only generate abstract images, which can not cover all scenarios in art design.

The generated covers are diverse and often quite well suited to musical compositions. Nevertheless, it is not always easy to interpret the connections taken into account by the generator for choosing the shape parameters. Sometimes, the covers of two completely different tracks may turn out to be visually similar. It is assumed that in order to obtain more stable and explicable patterns in the generation of covers, it is necessary to enlarge the dataset, as well as significantly complicate the generator architecture used. The basis of such a model can be an implemented recurrent generator, modified to transform a sequence of musical fragments embeddings into a sequence of figures, for example, using the transformer architecture~\cite{vaswani2017attention}. However, such developments require preliminary theoretical research in the field of vector image generation.

For the additional network engaged in the creating of captions, the main difficulty in practice is to ensure that the color is sufficiently contrasting with the background. In real covers, artists often use additional visual effects (shadows, contrasting strokes, glow, background fill) to achieve matching of slightly contrasting colors; a similar approach may be implemented for this solution in the future. 
Another difficult task is to wrap text inside the bounding box estimating the best number of lines. Currently, this is not provided and the text may not fit the canvas completely. In the future, it is supposed to use  the functionality of the HarfBuzz~\cite{harfbuzz} text generation library or the Pango~\cite{pango} rendering library.

\section{Conclusion}\label{sec:conclusion}
There exist music services offering musicians computer-generated images as cover images for their musical compositions; however, they do not consider the music track itself.
In this work, we have developed the CoverGAN model to generate vector images conditioned by a music track and its emotion. We have compared the model suggested to AttnGAN and DALL-E models for text-to-image generation.
The reported results prove that the proposed algorithm is competitive in the task of music cover synthesis. 
This is also good evidence that vector graphics synthesis is a promising research direction. 
%The main limitation of this approach is dictated by the set of parametric objects to be drawn on the image.  
The approach applied is limited only with the parametric space where the image is located.
In the future, it is planned to generate B\'ezier curves with an arbitrary number of segments based on an approach similar to Im2Vec~\cite{reddy2021im2vec}, as well as potentially some concrete shapes (natural scene, human silhouette).

\section{Acknowledgements}
This work was prepared at the expense of a grant for supporting research centers in the field of artificial intelligence, including in the field of "strong" artificial intelligence, trustworthy artificial intelligent systems, and ethical aspects of artificial intelligence application, provided by the Analytical Center for the Government of the Russian Federation in accordance with Subsidy agreement (subsidy agreement identifier 000000D730321P5Q0002) and agreement No. 70-2021-00141 from November 2, 2021.

{\small
\bibliographystyle{ieee_fullname}
\bibliography{PaperForReview}
}

\end{document}